\documentclass{JHEP3}

\usepackage{Pformel,Pmeta}
\usepackage{amsmath,amsfonts,amsthm,times}

\preprint{PTA/08-019}

\title{A family of solvable non-rational conformal field theories}

\author{Sylvain Ribault
\\ 
 Laboratoire de Physique Th\'eorique et Astroparticules, UMR5207 CNRS-UM2,
 \\
 Universit\'e Montpellier II, Place E. Bataillon,
 \\
 34095 Montpellier Cedex 05, France 
 \\
 {\footnotesize \tt ribault@lpta.univ-montp2.fr }
}

\abstract{We find non-rational conformal field theories in two dimensions, which are solvable due to their correlators being related to correlators of Liouville theory. Their symmetry algebra consists of the dimension-two stress-energy tensor, and two dimension-one fields. The theories come in a family with two parameters: the central charge $c$ and a complex number $m$. The special case $m=0$ corresponds to Liouville theory (plus two free bosons), and $m=1$ corresponds to the $\Hp$ model. In the case $m=2$ we show that the correlators obey third-order differential equations, which are associated to a subsingular vector of the symmetry algebra. }

\makeatletter\let\default@color\current@color\makeatother 

\begin{document}

\zeq\section{Introduction}

Conformal field theories in two dimensions have many applications in string theory and condensed-matter physics. In some applications the system under consideration, say strings in a given space-time, dictates the theory to be studied. In some other cases however, a CFT is used as a phenomelogical theory for a system without a known first-principles description. The relevant theory must then be chosen among the known CFTs according to the desired properties of the system. This approach relies on the existence of a large enough number of CFTs among which to choose, and on a good enough knowledge of their properties. Many rational CFTs are known, and their properties are often well-understood. However most CFTs are non-rational, and only a few non-rational CFTs have been solved so far, among them Liouville theory \cite{tes01a} and the $\Hp$ model \cite{gaw91}. It is therefore interesting to expand the small family of solvable non-rational CFTs. 

The inspiration for the present article comes from the relation \cite{rt05} between the correlation functions of the $\Hp$ model and correlation functions of Liouville theory. Namely, correlators of generic $\Hp$ primary fields are related to correlators involving both generic Liouville primary fields, and the particular field $V_{-\frac{1}{2b}}$ where $b$ is related to the central charge of Liouville theory by $c=1+6(b+b^{-1})^2$. This field is known to be a degenerate field with a null vector at level two, which implies that Liouville correlators involving this field obey second-order differential equations \cite{bpz84}. But there is an infinite series of degenerate fields in Liouville theory, which lead to higher-order differential equations. It is therefore natural to wonder whether Liouville correlators involving such degenerate fields -- or even generic fields -- can be related to correlators in some models which would be generalizations of the $\Hp$ model. 

We will propose a positive answer to this question, and sketch some properties of the resulting models. In contrast to the usual situation in the study of conformal field theory, we start with the solution of the models, namely an ansatz for their correlators in terms of Liouville correlators. The problem is then to reconstruct the models, and in particular their symmetry algebra, from their correlators. This is a priori a non-trivial problem, because our ansatz gives us only correlators of primary fields, but the notion of a primary field is not even defined if we do not know the symmetry algebra. Fortunately, the correct symmetry algebra will be suggested by the Lagrangian description of the models, which is very similar to the free-field description of the $\Hp$ model. (The new Lagrangians however do not correspond to sigma models.) We will then check that this symmetry algebra is consistent with the properties of the correlators, in particular in the case when they obey third-order differential equations. 

\zeq\section{An ansatz for the correlation functions}

Our ansatz will be directly inspired by the $\Hp$-Liouville relation, we will therefore begin by a reminder of this relation. More details and references can be found in \cite{rt05}. 

\subsection{Reminder of the $\Hp$-Liouville relation} 

In the conformal bootstrap approach, the $\Hp$ model is defined by its symmetry algebra, and primary fields which transform in a certain way with respect to this algebra. Solving the model means finding the correlation functions of the primary fields; these correlation functions are subject to symmetry and consistency conditions. Let us briefly review these objects.
The chiral symmetry algebra of the $\Hp$ model is the affine Lie algebra $\asl$, which is generated by three currents $J^3(z),J^+(z),J^-(z)$ with the following operator product expansions (OPEs):
\bea
J^3(z)J^3(w) &=& -\frac{\frac{k}{2}}{(z-w)^2} + O(1)\ ,
\label{jtjt}
\\
J^3(z)J^\pm(w)&=& \pm \frac{J^\pm(w)}{z-w} + O(1)\ ,
\label{jtjpm}
\\
J^+(z)J^-(w) &=& -\frac{k}{(z-w)^2}+ \frac{2J^3(w)}{z-w}+ O(1)\ ,
\label{jpjm}
\eea
where $k>2$ is called the level of the affine Lie algebra. As a consequence, the $\Hp$ model is a conformal field theory with central charge $c=\frac{3k}{k-2}$; the stress-energy tensor which generates the conformal transformations can be deduced from the currents by the Sugawara construction \cite{fms97}
\bea
T = \frac{1}{2(k-2)}\left[ (J^+J^-)+(J^-J^+)-2(J^3J^3)\right]\ ,
\label{tjj}
\eea
where we use the following definition of the normal-ordered product of operators:
\bea
(AB)(z) =\frac{1}{2\pi i}\oint_z dx\ \frac{dx}{x-z}A(x)B(z)  \ .
\label{abz}
\eea
Affine primary operators $\Phi^j(\mu|w)$ with spin $j$, isospin $\mu$ and worldsheet position $w$ are defined by their OPEs with the currents, from which their OPEs with the stress-energy tensor follows:
\bea
J^-(z) \Phi^j(\mu|w) &=& \frac{\mu}{z-w} \Phi^j(\mu|w) + O(1)\ ,
\label{jmph}
\\
J^3(z) \Phi^j(\mu|w) &=& \frac{\mu\pp{\mu} }{z-w} \Phi^j(\mu|w) +O(1)\ ,
\label{jtph}
\\
J^+(z) \Phi^j(\mu|w) &=& \frac{\mu\ppd{\mu} -\frac{j(j+1)}{\mu}}{z-w} 
\Phi^j(\mu|w) + O(1) \ ,
\label{jpph}
\\
T(z) \Phi^j(\mu|w) &=& \frac{\Delta_j}{(z-w)^2} \Phi^j(\mu|w) + \frac{1}{z-w}\pp{w} \Phi^j(\mu|w) +O(1) \ ,
\label{tph}
\eea
where the conformal dimension of the field $\Phi^j(\mu|w)$ is
\bea
\Delta_j = -\frac{j(j+1)}{k-2}\ .
\label{dj}
\eea
The fields $\Phi^j(\mu|w)$ have similar OPEs with the antiholomorphic currents $\bJ^a(\bz)$ and stress-energy tensor $\bar{T}(\bz)$, but their dependences on $\bz$ and $\bar{\mu}$ will be kept implicit.
The spin takes values $j\in -\frac12 +i\R$, corresponding to continuous representations of the algebra $\asl$. 

In the $\Hp$ model all fields are either affine primaries or affine descendents thereof, and correlators of affine descendents can be deduced from correlators of affine primaries by the Ward identities. Therefore, in order to solve the model on the Riemann sphere with coordinates $z,\bz$, it is enough to determine the $n$-point functions of affine primaries for all $n$:
\bea
\Omega_n \equiv \la \prod_{i=1}^n \Phi^{j_i}(\mu_i|z_i) \ra\ .
\label{omn}
\eea
Now all such correlators of affine primaries are related to Liouville theory correlators as follows:
\bea
\Omega_n = \delta^{(2)}(\tsum_{i=1}^n\mu_i) |u|^2 |\Theta_n|^{\frac{1}{b^2}} \la \prod_{i=1}^n V_{\al_i}(z_i) \prod_{a=1}^{n-2} V_{-\frac{1}{2b}}(y_a)\ra\ ,
\label{htli}
\eea
where we consider Liouville theory at parameter 
\bea
b = \frac{1}{\sqrt{k-2}} \ ,
\label{bk}
\eea
with primary operators $V_{\al}(z)$ with momenta
\bea
\al=b(j+1)+\frac{1}{2b}\ ,
\label{alj}
\eea
and conformal dimensions
\bea
\Delta_\al = \al(b+b^{-1}-\al)\ .
\label{dal}
\eea
The objects $u$ and $\Theta$ are defined by
\bea
u=\tsum_{i=1}^n \mu_iz_i \scs \Theta_n=\frac{\prod_{i<j} z_{ij} \prod_{a<b} y_{ab}}{\prod_{i=1}^n\prod_{a=1}^{n-2} (y_a-z_i)}\ ,
\label{uth}
\eea
while the positions $y_a$ of the operators $V_{-\frac{1}{2b}}(y_a)$ are determined in terms of $\mu_i,z_i$ by Sklyanin's change of variables
\bea
\sum_{i=1}^n \frac{\mu_i}{t-z_i} = u\frac{\prod_{a=1}^{n-2} (t-y_a)}{\prod_{i=1}^n (t-z_i)}\ .
\label{skl}
\eea

\subsection{Motivation of the ansatz}

Let us replace the fields $V_{-\frac{1}{2b}}$ in the $\Hp$-Liouville relation (\ref{htli}) with more general fields $V_{-\frac{m}{2b}}$ with $m\in \C$, and modify the prefactors as well:
\bea
\boxed{\Omega^{(m)}_n = \delta^{(2)}(\tsum_{i=1}^n\mu_i) |u|^{2\lambda} |\Theta_n|^{2\theta} \la \prod_{i=1}^n V_{\al_i}(z_i) \prod_{a=1}^{n-2} V_{-\frac{m}{2b}}(y_a)\ra}\ ,
\label{hmli}
\eea
where $\lambda$ and $\theta$ are numbers to be determined.
Can we still interpret $\Omega^{(m)}_n$ as an $n$-point function (\ref{omn}) of fields $\Phi^j(\mu|z)$ in some conformal field theory? In the process we keep the definitions of $u,\Theta$ and $y_a$ eqs. (\ref{uth},\ref{skl}), but the nature of the operator $\Phi^j(\mu|z)$ will be modified: it will now be a primary field with respect to an unknown symmetry algebra, with an unknown conformal dimension $\Delta^{(m)}_j$ with respect to a Virasoro algebra with an unknown central charge.

We hope that this interpretation of $\Omega^{(m)}_n$ exists for any choice of $m$, but we expect the values of the extra parameters $\lambda,\theta$ in (\ref{hmli}) to be determined in terms of $m$ by symmetry and other requirements. 
Let us first consider the constraints on the three-point function due to global conformal invariance. This 
requires that the dependence of $\Omega^{(m)}_3$ on the worldsheet coordinates $z_1,z_2,z_3$ should be
\bea
\Omega^{(m)}_3 \propto |z_{12}|^{-2\left(\Delta^{(m)}_{j_1}+\Delta^{(m)}_{j_2} -\Delta^{(m)}_{j_3}\right)} |z_{13}|^{-2\left(\Delta^{(m)}_{j_1}+\Delta^{(m)}_{j_3} -\Delta^{(m)}_{j_2}\right)} 
|z_{23}|^{-2\left(\Delta^{(m)}_{j_2}+\Delta^{(m)}_{j_3} -\Delta^{(m)}_{j_1}\right)} 
\ ,
\label{ommz}
\eea
with an arbitrary prefactor depending on $\mu_1,\mu_2,\mu_3$.
According to our ansatz, $\Omega^{(m)}_3$ is actually related to a four-point Liouville correlator, which behaves as 
\begin{multline}
\la V_{\al_1}(z_1)V_{\al_2}(z_2)V_{\al_3}(z_3) V_{-\frac{m}{2b}}(y_1)\ra 
\\ = |z_{12}|^{2\Delta_{-\frac{m}{2b}}-2\Delta_{12}} |z_{13}|^{-2\Delta_{-\frac{m}{2b}}-2\Delta_{13}} |z_{23}|^{2\Delta_{-\frac{m}{2b}}-2\Delta_{23}} |y_1-z_2|^{-4\Delta_{-\frac{m}{2b}}} {\cal G}(z)\ ,
\label{vvvv}
\end{multline}
where we use the notations $\Delta_{12}=\Delta_{\al_1}+\Delta_{\al_2}-\Delta_{\al_3}$ for the Liouville conformal dimensions, and ${\cal G}(z)$ for a known function of the cross-ratio $z=\frac{z_{32}(y-z_1)}{z_{31}(y-z_2)}$ whose precise form is not needed here. From the change of variables (\ref{skl}) we have $u(y-z_2) = \mu_2z_{12}z_{23}$ (in the limit $t\rar z_2$), 
and in addition $z=1+\frac{\mu_3}{\mu_2}$ actually does not depend on $z_1,z_2,z_3$. Thus, the equations (\ref{hmli}) and (\ref{vvvv}) imply
\bea
\Omega^{(m)}_3 \propto |u|^{2\lambda+6\theta+4\Delta_{-\frac{m}{2b}}} |z_{12}|^{-2\theta-2\Delta_{-\frac{m}{2b}} -2\Delta_{12}} |z_{13}|^{-2\theta-2\Delta_{-\frac{m}{2b}} -2\Delta_{13}} |z_{23}|^{-2\theta-2\Delta_{-\frac{m}{2b}} -2\Delta_{23}}\ .
\eea
Comparing with the condition (\ref{ommz}) reveals that global conformal invariance requires
\bea
\lambda+3\theta+2\Delta_{-\frac{m}{2b}} = 0 \scs \Delta^{(m)}_{j_i} =\Delta_{\al_i}+\Delta_{-\frac{m}{2b}}+\theta\ .
\label{ltd}
\eea
In order to derive the extra relation which we need for fully determining $\lambda,\theta$ and $\Delta_{j_i}^{(m)}$, we will use a more heuristic reasoning. Consider the behaviour of our ansatz (\ref{hmli}) near $y_a=y_b$. In the $\Hp$ case ($m=1$), correlators are continuous at such points, a property which is particularly important in the study of the $\Hp$ model on the disc \cite{hr06}. More generally, in the cases when the Liouville field $V_{-\frac{m}{2b}}$ is degenerate, the leading behaviour of two such fields when coming close is 
\bea
V_{-\frac{m}{2b}}(y_a) V_{-\frac{m}{2b}}(y_b) \underset{y_{ab}\rar 0}{\sim} |y_{ab}|^{-4\Delta_{-\frac{m}{2b}}+2\Delta_{-\frac{m}{b}}} V_{-\frac{m}{b}}(y_a)\ ,
\eea
and the behaviour of our ansatz is thus 
\bea
\Omega_n^{(m)} \underset{y_{ab}\rar 0}{\sim} |y_{ab}|^{-4\Delta_{-\frac{m}{2b}}+2\Delta_{-\frac{m}{b}} +2\theta} \ .
\eea
Assuming the critical exponent $-4\Delta_{-\frac{m}{2b}}+2\Delta_{-\frac{m}{b}} +2\theta$ to vanish, and solving eq. (\ref{ltd}) as well, we obtain 
\bea
\boxed{
\theta =\frac{m^2}{2b^2} \scs \lambda =m(1+b^{-2}(1-m)) \scs \Delta_{j}^{(m)}=\Delta_{\al} -\frac{m}{4}(2+2b^{-2}-b^{-2}m)}\ .
\label{tele}
\eea
These relations will be assumed to hold not only when the Liouville field $V_{-\frac{m}{2b}}$ is degenerate, but in all cases. Notice that it is possible to define a ``spin'' variable $j$ such that 
\bea
\Delta^{(m)}_j=-(j+1)(b^2j+m-1) \ ,
\label{dmj}
\eea
in which case the momentum $\al$ of the corresponding Liouville field $V_\al(z)$ is 
\bea
\al= b(j+1)+\frac{m}{2b} \ .
\label{aljm}
\eea

In the following we will argue that the ansatz (\ref{hmli}), together with the particular values (\ref{tele}) for the parameters $\lambda,\theta$ and the conformal weights of the fields $\Phi^j(\mu|z)$, has a meaning in terms of correlators in a conformal field theory.

\zeq\section{Symmetry algebra of the new theories}

We have checked that our ansatz is consistent with global conformal invariance, let us now study the rest of the conformal symmetry, and the possible extra symmetries which underlie our ansatz. To do this, it will be convenient to introduce a Lagrangian description of the corresponding theories. We will therefore propose such a description, check that it reproduces the relation (\ref{hmli}) with Liouville theory, and then use it for deriving the symmetries.

\subsection{Lagrangian description}

We use the same bosonic fields $(\phi,\beta,\g)$ as in the $\Hp$ model \cite{hos00}, and we propose the following Lagrangian and stress-energy tensor:
\bea
{\cal L}^{(m)} &=& \frac{1}{2\pi} \left[\p\phi\bp\phi +\beta\bp\g+\bar{\beta}\p\bg +b^2(-\beta\bar{\be})^m e^{2b\phi}\right]\ ,
\label{lm}
\\
T^{(m)} &=& -\be\p\g-(\p\phi)^2+(b+b^{-1}(1-m))\p^2\phi\ ,
\label{tm}
\eea
where $-\beta\bar{\beta}$ is assumed to be real positive.

The interaction term ${\cal L}^{(m)}_{int} = (-\beta\bar{\be})^m e^{2b\phi}$ is marginal with respect to the stress-energy tensor $T^{(m)}$. In the cases $m=0$ (Liouville theory) and $m=1$ ($\Hp$ model) it is known to be exactly marginal. This follows from the exponential dependence of the interaction term on $\phi$, a feature which is still present for arbitrary values of $m$. In perturbative calculations, many terms in the series expansion of $\exp -\int {\cal L}^{(m)}_{int}$ will then yield vanishing contributions due to $\phi$-momentum conservation in the free theory. We therefore expect the interaction to be exactly marginal, which
implies that the corresponding theories have conformal symmetry.

In the case $m=1$ of the $\Hp$ model, integrating out the non-dynamical fields $\beta,\bar{\be}$ in the path integral $\int {\cal D}\phi{\cal D}\beta{\cal D}\bar{\beta}{\cal D}\g{\cal D}\bg\  e^{-\int d^2w\ {\cal L}^{(m)}} $ yields a sigma model, whose target space is indeed $\Hp$. In the general case, it is still possible to integrate out $\beta,\bar{\be}$ if $\Re m>\frac12$, but the resulting Lagrangian has no sigma model interpretation because it involves higher powers of $\bp\g\p\bg$.

The relation between $\Hp$ and Liouville correlators has been rederived by Hikida and Schomerus using a path-integral computation \cite{hs07}. We will now emulate this calculation in order to rederive our ansatz (\ref{hmli}) from the path-integral definition of a theory with the Lagrangian (\ref{lm}). Of course, this path-integral definition will be complete only after we specify the fields $\Phi^j(\mu|z)$ in terms of $(\phi,\beta,\g)$:
\bea
\Phi^j(\mu|z)=|\mu|^{2m(j+1)} e^{\mu\g-\bar{\mu}\bg} e^{2b(j+1)\phi}\ .
\label{pjmz}
\eea
The path-integral definition of the $n$-point function is then
\bea
\Omega_n^{(m)} = \int {\cal D}\phi{\cal D}\beta{\cal D}\bar{\beta}{\cal D}\g{\cal D}\bg\ \ e^{-\int d^2w\ {\cal L}^{(m)}} \prod_{i=1}^n \Phi^{j_i}(\mu_i|z_i)\ .
\label{ommp}
\eea
Let us perform the integrations over $\g$ and $\bg$, and then $\beta$ and $\bar{\beta}$:
\bea
\Omega_n^{(m)} = \delta^{(2)}(\tsum_{i=1}^n\mu_i) \int {\cal D}\phi\  e^{-\frac{1}{2\pi}\int d^2w \left(\p\phi\bp \phi +b^2 \left|u\sum_{i} \frac{\mu_i}{w-z_i} \right|^{2m} e^{2b\phi} \right) } \prod_{i=1}^n |\mu_i|^{2m(j_i+1)} e^{2b(j_i+1)\phi}\ ,
\eea
where we still have $u=\sum_{i=1}^n \mu_iz_i$. We then perform the change of integration variable
\bea
\phi \rar \phi-mb^{-1}\log|u|\ .
\eea
This produces a global factor $|u|^{2m(1+b^{-2}(1-m))}$, due to an implicit worldsheet curvature term in the Lagrangian, which corresponds to the linear dilaton term $(b+b^{-1}(1-m))\p^2\phi$ in the stress-energy tensor $T^{(m)}$, eq. (\ref{tm}). (For simplicity we have used flat space as a model of the Riemann sphere and omitted the worldsheet curvature term. This subtlety is dealt with in \cite{hs07}.) Then, we define $y_a$ as the zeroes of $\sum_{i=1}^n\frac{\mu_i}{w-z_i}$ as in eq. (\ref{skl}), and perform the change of integration variable
\bea
\varphi(w)=\phi(w)+mb^{-1}\log\left|\frac{\prod_{a=1}^{n-2} (w-y_a)}{\prod_{i=1}^n (w-z_i)}\right|\ .
\eea
This yields
\begin{multline}
\Omega_n^{(m)} = \delta^{(2)}(\tsum_{i=1}^n\mu_i)\ |u|^{2m(1+b^{-2}(1-m))}\ |\Theta_n|^{\frac{m^2}{b^2}} 
\\ \times
\int {\cal D}\varphi\ e^{-\frac{1}{2\pi}\int d^2w\left(\p\varphi\bp\varphi +b^2 e^{2b\varphi}\right)} \prod_{i=1}^n e^{(2b(j_i+1)+\frac{m}{b})\varphi(z_i)} \prod_{a=1}^{n-2} e^{-\frac{m}{b}\varphi(y_a)}\ ,
\end{multline}
where $\Theta_n$ was defined in eq. (\ref{uth}). The second line of this formula is the path-integral version of the Liouville theory correlator which appears in our ansatz (\ref{hmli}), with the expected values for the Liouville momenta $\al_i=b(j_i+1)+\frac{m}{2b}$. And the prefactors also agree with the expectations (\ref{tele}). 

Therefore, the path-integral calculation provides a Lagrangian definition (\ref{lm}) for the new theories whose correlators we conjectured. This definition will allow us to easily study the symmetries of these theories.

\subsection{Symmetry algebra}

Let us interpret the theories with Lagrangian (\ref{lm}) as free theories of the fields $(\phi,\beta,\g,\bar{\beta},\bg)$ with contractions
\bea
\la \phi(z)\phi(w) \ra = -\log |z-w| \scs \la \beta(z)\g(w) \ra=\frac{1}{w-z}\ ,
\eea
deformed by the interaction term ${\cal L}^{(m)}_{int}=(-\beta\bar{\beta})^m e^{2b\phi}$. The chiral symmetry algebra is then the set of holomorphic fields whose OPEs with ${\cal L}^{(m)}_{int}$ vanish up to total derivatives. Such chiral fields can be constructed from the basic holomorphic fields $\p\phi,\beta,\g$ by using the normal-ordered product (\ref{abz}), see for instance \cite{bs92} for the rules of computing with this product. 

We already know one chiral field, namely the stress-energy tensor $T^{(m)}$ (\ref{tm}), which generates a Virasoro algebra with central charge
\bea
c^{(m)} = 3+6(b+b^{-1}(1-m))^2\ .
\label{cm}
\eea
But in the case of the $\Hp$ model ($m=1$) we know that the creation modes of this Virasoro algebra are not enough for generating the full spectrum of the model from the fields $\Phi^j(\mu|z)$. These fields are indeed affine primaries, that is primary fields with respect to the much larger symmetry algebra $\asl$. Can we find a larger algebra for general values of $m$? Let us introduce the following holomorphic currents:
\bea
J^-&=&\beta\ ,
\label{jm}
\\
J^3&=&-\beta\g -mb^{-1}\p\phi\ ,
\label{jt}
\\
J^+&=&\beta\g^2+2mb^{-1}\g\p\phi-(m^2b^{-2}+2)\p\g\ ,
\label{jp}
\eea
where normal ordering is implicitly assumed when needed. The OPEs of these currents obey the relations
(\ref{jtjt}-\ref{jpjm}) of an $\asl$ algebra at level $k=2+m^2b^{-2}$. However, only the currents $J^-$ and $J^3$ are symmetries of our model. The current $J^+$ indeed has a nontrivial OPE with ${\cal L}^{(m)}_{int}$, and is no symmetry. And the stress-energy tensor which we could build from $J^-,J^3,J^+$ by the Sugawara construction (\ref{tjj}) is therefore also no symmetry. It actually differs from the
stress-energy tensor (\ref{tm}), with respect to which the interaction ${\cal L}^{(m)}_{int}$ is marginal. The chiral symmetry algebra is therefore generated by the three fields $T,J^3,J^-$ (where from now on we omit the superscipt of $T^{(m)}$), and we compute their OPEs as
\bea
T(z)T(w) &=& \frac{\frac12 c^{(m)}}{(z-w)^4} + \frac{2T(w)}{(z-w)^2} + \frac{\p T(w)}{z-w} + O(1)\ ,
\label{tt}
\\
T(z) J^-(w) &=& \frac{J^-(w)}{(z-w)^2} +\frac{\p J^-(w)}{z-w} +O(1)\ ,
\\
T(z) J^3(w) &=&\frac{(1-m)(1-mb^{-2})}{(z-w)^3}+
\frac{J^3(w)}{(z-w)^2} +\frac{\p J^3(w)}{z-w} +O(1)\ ,
\label{tjt}
\\
J^-(z)J^-(w)&=& O(1)\ ,
\\
J^3(z)J^-(w) &=& \frac{J^-(w)}{w-z}+O(1)\ ,
\\ 
J^3(z) J^3(w) &= &- \frac{1+\frac12 m^2b^{-2}}{(z-w)^2} +O(1)\ .
\label{jzjw}
\eea
This looks very much like the subalgebra of the affine algebra $\asl$ obtained by removing the current $J^+$, except that there is now a central term in the $TJ^3$ OPE, so that $J^3$ is no longer a primary field. Notice that this algebra, unlike the Lagrangian ${\cal L}^{(m)}$, is invariant under the duality
\bea
b\rar b^{-1}\scs m\rar mb^{-2}\ .
\eea
Notice also that setting $J^-=0$ yields a smaller algebra generated by $T$ and $J^3$, which might be interesting as well.\footnote{We could even consider a three-parameter family of $T,J^3$ algebras with arbitrary central terms $T(z)T(w) = \frac{\frac12 c}{(z-w)^4}+\cdots,\ T(z)J^3(w)=\frac{Q}{(z-w)^3}+\cdots $ and $ J^3(z)J^3(w)=-\frac{\frac{k}{2}}{(z-w)^2}+\cdots$. The relation $c=1-6\frac{Q^2}{k}$ defines a subfamily where the identification $T=-\frac{1}{k}(J^3J^3)+\frac{Q}{k}\p J^3$ is allowed. After introducing a boson $\varphi$ such that $J^3=\p\varphi$ this corresponds to a linear dilaton theory. Our parameters $m,b$ parametrize a different subfamily.}

It can be checked that the fields $\Phi^j(\mu|z)$ are primary with respect to our chiral algebra. We can indeed compute the OPEs of their free-field realization (\ref{pjmz}) with $J^-,J^3$ and $T$, which respectively reproduces the OPEs (\ref{jmph}), (\ref{jtph}) and (\ref{tph}) with however the conformal dimension $\Delta_j^{(m)}$ (\ref{dmj}). And we can check that the correlators (\ref{hmli}) have the correct behaviour under global symmetry transformations. We already performed this analysis in the case of the global conformal transformations of the three-point function $\Omega^{(m)}_3$, this is how we found $\Delta_j^{(m)}$ in the first place. The global Ward identity for $J^-$ is $(\sum_{i=1}^n \mu_i) \Omega_n^{(m)}=0$, which is obviously satisfied. We furthermore compute $\left(\sum_{i=1}^n \mu_i\pp{\mu_i} \right) \Omega_n^{(m)} = -(1-m)(1-mb^{-2}) \Omega_n^{(m)}$. This is what is expected knowing that $J^3$ is no longer primary, see the $TJ^3$ OPE (\ref{tjt}).

These simple consistency checks leave two questions about our determination of the symmetry algebra:
\begin{enumerate}
 \item {\bf Is the symmetry algebra large enough?} We still have to show that the descendent fields obtained by repeatedly acting on the primaries $\Phi^j(\mu|z)$ with the creation modes of $T,J^3,J^-$ do span the spectrum. In particular, can the $\Phi^{j_1}(\mu_1|z_1)\Phi^{j_2}(\mu_2|z_2)$ OPE be written as a sum over such descendents? In principle, we could address this issue by studying the $z_{12}\rar 0$ expansion of our ansatz (\ref{hmli}) for the correlators $\Omega_n^{(m)}$. This would however be tedious. Instead, let us focus on the $m=1$ case. The spectrum of the $\Hp$ model is known to be generated by the modes of $J^3,J^+,J^-$, but we will now argue that the modes of $T,J^3,J^-$ are actually enough. Let us define these modes by $T(z)=\sum_{n\in\Z} L_n z^{-n-2}$ and $J^a(z)=\sum_{n\in\Z} J^a_n z^{-n-1}$; creation modes are $J^a_{n<0}$ and $L_{n<0}$, while primary fields $\Phi^j(\mu|z)$ correspond to primary states $|p\rangle$ such that $L_{n>0}|p\rangle = J^a_{n>0}|p\rangle=0$.
 Any level one $J^+$-descendent state can by definition be written as $|d\rangle= J^+_{-1}|p'\rangle$ with $|p'\rangle$ an affine primary state in the $\Hp$ spectrum. The spectrum being made of continuous representations, this primary can be rewritten as $|p'\rangle = J_0^-|p\rangle$ with $|p\rangle$ another primary. Using the Sugawara construction (\ref{tjj}), we then have
 \bea
 |d\rangle =  J^+_{-1}J_0^-|p\rangle =\left[-J^-_{-1}J^+_0 +2J^3_{-1}J^3_0 +(k-2)L_{-1}\right] |p\rangle\ ,
 \label{dp}
 \eea
 which is manifestly a combination of $L_{-1},J^3_{-1},J^-_{-1}$-descendents of affine primary states. This reasoning can be iterated to higher level $J^+$-descendent states. This proof is special to $m=1$ and cannot be generalized, but it demonstrates that our $T,J^3,J^-$ symmetry algebra is likely large enough.
 
 \item {\bf Are the chiral fields algebraically independent?} In the $\Hp$ model, the fact that the chiral fields $T,J^+,J^3,J^-$ are not independent but related by the Sugawara construction implies that the correlators obey Knizhnik--Zamolodchikov differential equations. In our generalized theories with parameter $m$, correlators (\ref{hmli}) do obey differential equations for values of $m$ such that the field $V_{-\frac{m}{2b}}$ is a Liouville degenerate field. This happens if
 \bea
 m = p + b^2 q \scs p,q=0,1,2\cdots
 \label{mpq}
 \eea
 We therefore expect that the structure of our chiral algebra becomes in some sense reducible for these values of $m$, so that differential equations can be derived for the correlators. This is what we will  explicitly demonstrate in the case $m=2$.
\end{enumerate}

\zeq\section{Differential equation in the case $m=2$}

We will first derive the third-order differential equations satisfied by the correlator $\Omega^{(2)}_n$ from the Belavin--Polyakov--Zamolodchikov equations satisfied by the corresponding Liouville correlators, and then check that these equations can be recovered from our symmetry algebra.

\subsection{Third-order BPZ equation}

The Virasoro module generated by the Liouville field $V_{-\frac{1}{b}}$ has a null vector at level three \cite{bpz84}:
\bea
\chi_3 = \left[(-1+2b^{-2}) L_{-3} +2L_{-1}L_{-2}+\frac12 b^2 L_{-1}^3\right] V_{-\frac{1}{b}}\ .
\label{chit}
\eea
Assuming that this null vector vanishes implies that correlators involving $V_{-\frac{1}{b}}$ obey third-order BPZ differential equations. In particular, let us write the BPZ equation associated to a degenerate field $V_{-\frac{1}{b}}(y)$ in the Liouville correlator which appears in our ansatz (\ref{hmli}). We call $y_b\neq y$ the insertion points of the other degenerate fields so that $y_a=(y,y_b)$; moreover we call $z_I=(z_i,y_b)$ the positions of all fields except  $V_{-\frac{1}{b}}(y)$. The BPZ equation then involves the following differential operator:
\bea
{\cal D} = \frac{b^2}{2} \frac{\p^3}{\p y^3} 
+ \sum_I\left[\pp{y}  \frac{2}{y-z_I}\pp{z_I} + \frac{2\Delta_{\al_I}}{(y-z_I)^2}\pp{y} \right]
+ \sum_I \left[\frac{\Delta_{-\frac{1}{b}}+2}{(y-z_I)^2}\pp{z_I}+ \frac{2\Delta_{-\frac{1}{b}}\Delta_{\al_I}}{(y-z_I)^3}\right]  \ .
\eea
This BPZ equation implies ${\cal D}' \Omega_n^{(2)}= 0$ where ${\cal D}'=\Theta_n^\frac{2}{b^2} {\cal D} \Theta_n^{-\frac{2}{b^2}}$, with $\Theta_n$ defined by eq. (\ref{uth}). We thus wish to compute ${\cal D}'$; at the same time we should perform the change of variables $(z_i,y_a)\rar (z_i,\mu_i)$ defined by eq. (\ref{skl}). In particular we should rewrite $\pp{z_i}=\left.\pp{z_i}\right|_{y_a} $ in terms of $\frac{\delta}{\delta z_i} \equiv \left.\pp{z_i}\right|_{\mu_j}$. This will be done thanks to the identity \cite{rt05}
\bea
\sum_{i=1}^n\frac{1}{y-z_i}\pp{z_i} +\sum_b \frac{1}{y-y_b} \pp{y_b} -X\pp{y} = \sum_{i=1}^n \frac{1}{y-z_i}\frac{\delta}{\delta z_i}\ , 
\label{dzeq}
\eea
where we defined
\bea
X \equiv \sum_b\frac{1}{y-y_b}-\sum_{i=1}^n \frac{1}{y-z_i} = -\frac{\sum_{i=1}^n \frac{\mu_i}{(y-z_i)^3}}{\sum_{i=1}^n \frac{\mu_i}{(y-z_i)^2}} \ .
\label{xdef}
\eea
We also use 
\bea
\pp{y} = \sum_{i=1}^n \frac{\mu_i}{y-z_i} \pp{\mu_i}\ .
\label{ppy}
\eea
Explicit calculations yield
\begin{multline}
{\cal D}'= \frac{b^2}{2} \frac{\p^3}{\p y^3} -X\ppd{y} +2 \pp{y} L_{-2}(y) -4b^{-2} X L_{-2}(y) 
\\
+(2b^{-2}-1)\left[L_{-3}(y)-\sum_i \frac{\mu_i}{(y-z_i)^3}\pp{\mu_i} -X \sum_i\frac{\mu_i}{(y-z_i)^2}\pp{\mu_i}\right]\ ,
\label{tdt}
\end{multline}
where we defined
\bea
L_{-2}(y) &=& \sum_i\frac{1}{y-z_i}\left(\frac{\delta}{\delta z_i} +\frac{\Delta_{j_i}^{(2)}}{y-z_i}\right)\ ,
\label{lmdy}
\\
L_{-3}(y)&=&-\sum_i\frac{1}{(y-z_i)^2}\left(\frac{\delta}{\delta z_i} +\frac{2\Delta_{j_i}^{(2)}}{y-z_i}\right) \ ,
\label{lmty}
\eea
where the conformal dimensions $\Delta_{j_i}^{(2)}$ and $\Delta_{\al_i}$ are related by eq. (\ref{tele}).

The operator ${\cal D}'$ can be understood as a generalization of the $s\ell_2$ Knizhnik--Zamolodchikov differential operator, which in our notations can be written as \cite{rt05}
\bea
{\cal D}^{KZ} = b^2 \ppd{y} + L_{-2}(y)\ .
\eea

\subsection{Reformulation of the BPZ equation in terms of symmetry generators}

If our identification of the symmetry algebra is correct, the differential equation ${\cal D}'\Omega^{(2)}_n=0$ should have a reformulation in terms of the chiral fields $T,J^3,J^-$. This is a non-trivial requirement on ${\cal D}'$, and we will now show that it is satisfied. 

Let us denote $\la {\cal Q}\ra =\la \prod \Phi^{j_i}(\mu_i|z_i)\ra$. The actions of differential operators $\pp{y} $, $L_{-2}(y)$ and $L_{-3}(y)$ on $\la {\cal Q}\ra$ have immediate interpretations as insertions of the chiral fields $J^3(y),T(y)$ and $\p T(y)$, for instance
\bea
\pp{y} \la {\cal Q}\ra = \la \sum_i\oint_{z_i}dt \frac{J^3(t)}{y-t} {\cal Q} \ra = \la \oint_y dt\frac{J^3(t)}{t-y} {\cal Q} \ra = \la J^3(y) {\cal Q}\ra \ ,
\eea
where we used the formula (\ref{ppy}) for $\pp{y} $ and the OPE (\ref{jtph}) of $J^3$ with $\Phi^j(\mu|z)$. Iterating the actions of such differential operators yields results like
\bea
 \pp{y} L_{-2}(y) \la {\cal Q}\ra = \pp{y}\la T(y) {\cal Q} \ra = \la \p T {\cal Q}\ra +\la (J^3T)(y) {\cal Q}\ra
\ ,
\eea
where the normal-ordered product $(J^3T)$ is defined as previously, see eq. (\ref{abz}). Moreover, we have
\bea
\sum_i\frac{\mu_i}{(y-z_i)^3} \pp{\mu_i} \la {\cal Q}\ra = \frac12 \la \sum_i \oint_{z_i}dt \frac{\p^2J^3(t)}{y-t} {\cal Q}\ra = \la \tfrac12\p^2J^3(y){\cal Q}\ra\ .
\eea
Finally, the factors of $X$ in ${\cal D}'$ can be related to insertions of $\p J^- $ and $\p^2 J^-$, thanks to eq. (\ref{xdef}) and identities of the type
\bea
\sum_i\frac{\mu_i}{(y-z_i)^3} \la {\cal P}(y) {\cal Q}\ra = \frac12 \la \oint_y dt\frac{\p^2J^-(t)}{t-y} {\cal P}(y) {\cal Q}\ra = \la \tfrac12 (\p^2J^- {\cal P})(y)\ {\cal Q} \ra\ ,
\eea
which is valid for any operator ${\cal P}$.

Therefore, the equation ${\cal D}'\Omega_n^{(2)}=0$ with the operator ${\cal D}'$ given in eq. (\ref{tdt}) can be rewritten as $\la {\cal R}(y) \prod \Phi^{j_i}(\mu_i|z_i)\ra=0$, where 
\begin{multline}
{\cal R}=\tfrac12 b^2 (\p J^-(J^3(J^3J^3)))+2 (\p J^-( J^3T))+[2b^{-2}+1] (\p J^-\p T) 
\\
+\tfrac32 b^2 (\p J^-(J^3\p J^3))+ \tfrac12 [b^2+1-2b^{-2}] (\p J^- \p^2J^3)
\\
-\tfrac12 (\p^2 J^-(J^3J^3))+[-1+b^{-2}](\p^2 J^-\p J^3)-2b^{-2}(\p^2J^-T)\ .
\label{rop}
\end{multline}
We can already conjecture that, for all the values (\ref{mpq}) of $m$ such that the fields $V_{-\frac{m}{2b}}$ is degenerate, the BPZ equations can similarly be rewritten in terms of operators $T,J^-,J^3$. It is however not clear how to deduce our operator ${\cal R}$ from a null vector like (\ref{chit}), without performing the explicit calculations as we did. 

The equation $\la {\cal R}(y) \prod \Phi^{j_i}(\mu_i|z_i)\ra=0$ of course does not mean that the operator ${\cal R}$ should be set to zero, because this equation is valid only at special points $y=y_1\cdots y_{n-2}$. These points were defined (\ref{skl}) as the zeroes of $\varphi(t)=\sum_{i=1}^n\frac{\mu_i}{t-z_i}$; they can be characterized in terms of the operator $J^-$ by
\bea
\la J^-(y) \prod_{i=1}^n \Phi^{j_i}(\mu_i|z_i) \ra = 0\ .
\eea
Actually, for any operator ${\cal P}$ we have $\la (J^-{\cal P})(y) \prod_{i=1}^n \Phi^{j_i}(\mu_i|z_i) \ra = 0$. Therefore, the operator ${\cal R}$ is expected to vanish modulo operators of the type $(J^-{\cal P})$. 

In other words, we expect that this operator corresponds to a subsingular vector ${\cal R}|0\rangle$ in the vacuum module of our symmetry algebra. That is, if we would set the singular vector $J^-|0\rangle$ to zero, then ${\cal R}|0\rangle$ would become a singular vector in the resulting coset module. We are actually not setting $J^-|0\rangle$ to zero, but the presumptive subsingular vector ${\cal R}|0\rangle$ is nevertheless associated to differential equations satisfied by the correlation functions. Relations between subsingular vectors and differential equations were found previously by Dobrev \cite{dob95} in the context of finite-dimensional symmetry algebras, but they do not seem widespread in conformal field theory so far. 
\footnote{I am very grateful to Vladimir Dobrev for pointing out that the notion of a subsingular vector is relevant here, and for patiently explaning some of the literature on this topic to me.} 

\subsection{Subsingular vectors of the symmetry algebra}

In order to show that the third-order differential equation can be deduced from our symmetry algebra, we still have to prove that the operator ${\cal R}$ (\ref{rop}) corresponds to a subsingular vector. We will actually investigate the more general operator
\begin{multline}
{\cal R}_{\{\lambda_i\} } = \lambda_1 (\p J^-J^3J^3J^3) + \lambda_2(\p J^-J^3T) + \lambda_3(\p J^-\p T) + \lambda_4 (\p J^-J^3\p J^3) + \lambda_5(\p J^-\p^2J^3)
 \\
  + \lambda_6 (\p^2J^-J^3J^3)+ \lambda_7 (\p^2J^- \p J^3) + \lambda_8(\p^2J^-T)\ ,
\label{rli}
\end{multline}
which depends on arbitrary coefficients $\lambda_1\cdots \lambda_8$.
Here and in the following we use the shorthand notation $ABCD=(ABCD)=(A(B(CD)))$ for multiple normal orderings, see \cite{bs92} for more details on calculations involving such expressions.

We will perform this investigation with the help of the free fields $(\phi,\beta,\g)$. Since the map from $(J^-,J^3,T)$ to $(\phi,\beta,\g)$ defined by the equations (\ref{jm}), (\ref{jt}) and (\ref{tm}) is a morphism of algebras, it can indeed be used for determining whether ${\cal R}_{\{\lambda_i\} }$ generates a nontrivial ideal of the coset algebra obtained by modding out the ideal generated by $J^-$. (This is the algebraic formulation of our subsingular vector problem.) 
So let us compute the operator ${\cal R}_{\{\lambda_i\} }$ in terms of the fields $(\phi,\beta,\g)$, modulo operators of the type $(J^-{\cal P})=(\beta {\cal P})$. We will denote by $\simeq$ the equality of operators modulo such terms. For example, we find
\bea
(J^3(J^3J^3))&\simeq& 2\p\be\p\g-\tfrac12\p^2\be\g-3mb^{-1}\p\be\g\p\phi-m^3b^{-3}\p\phi^3\ ,
\\
(J^3T)&\simeq& \p\be\p\g+mb^{-1}\p\phi^3-m(1+b^{-2}(1-m))\p\phi\p^2\phi\ ,
\\
(J^3\p J^3)&\simeq& \tfrac12\p^2\be\g+mb^{-1}\p\be\g\p\phi+m^2b^{-2}\p\phi\p^2\phi\ ,
\\
(\p T) &\simeq& -\p\be\p\g-2\p\phi\p^2\phi+(b+b^{-1}(1-m))\p^3\phi\ ,
\\
(\p^2J^3)&\simeq & -\p^2\be\g-2\p\be\p\g-mb^{-1}\p^3\phi\ .
\eea
We then compute
\begin{multline}
{\cal R}_{\{\lambda_i\} }\simeq\left[2\lambda_1+\lambda_2-\lambda_3-2\lambda_5\right] \p\beta^2\p\g +\left[-3mb^{-1}\lambda_1+mb^{-1}\lambda_4\right] \p\be^2\g\p\phi 
\\
+\left[-m^3b^{-3}\lambda_1+mb^{-1}\lambda_2\right] \p\be\p\phi^3 
+\left[-m(1+b^{-2}(1-m))\lambda_2-2\lambda_3+m^2b^{-2}\lambda_4\right] \p\be\p\phi\p^2\phi 
\\
+\left[(b+b^{-1}(1-m))\lambda_3-mb^{-1}\lambda_5\right]\p\be\p^3\phi 
+\left[-\tfrac12\lambda_1+\tfrac12\lambda_4-\lambda_5+\lambda_6-\lambda_7\right]\p\be\p^2\be\g 
\\
+\left[m^2b^{-2}\lambda_6-\lambda_8\right]\p^2\be\p\phi^2
+\left[-mb^{-1}\lambda_7+(b+b^{-1}(1-m))\lambda_8\right]\p^2\be\p^2\phi\ .
\end{multline}
Let us solve the equation ${\cal R}_{\{\lambda_i\} }\simeq 0$, which when satisfied implies that ${\cal R}_{\{\lambda_i\} }$ corresponds to a subsingular vector. This equation leads to a system of 8 linear equations for the 8 unknowns $\lambda_1\cdots \lambda_8$. Actually, the first five equations form a closed subsystem of equations for $\lambda_1\cdots \lambda_5$. This subsystem has a nonzero solution only if its determinant vanishes, that is if 
\bea
(m-1)(m-2)(m-b^2)(m-2b^2)=0\ .
\eea
The four solutions of this equation correspond to the Liouville operators $V_{-\frac{m}{2b}}$ being degenerate, leading to BPZ differential equations of order 2 or 3. This is therefore a strong check of our claim that the ansatz (\ref{hmli}) does define conformal field theories with symmetry algebras generated by $T,J^3,J^-$. Let us perform a more detailed check in the case $m=2$, by explicitly computing the nonzero solution to the system ${\cal R}_{\{\lambda_i\} }=0$:
\bea
\begin{array}{c}
\lambda_1=\tfrac12 b^2\scs \lambda_2=2\scs \lambda_3=2b^{-2}+1 \scs \lambda_4=\tfrac32 b^2 \scs \lambda_5 = \tfrac12(b^2+1-2b^{-2}) \ , 
\vspace{2mm}
\\
\lambda_6=-\tfrac12 \scs \lambda_7=-1+b^{-2} \scs \lambda_8=-2b^{-2}\ .
\end{array}
\eea
With these values of of $\lambda_i$, the operator ${\cal R}_{\{\lambda_i\} }$ (\ref{rli}) does agree with the operator ${\cal R}$ (\ref{rop}), which we found by reformulating the third-order BPZ equation.  

\zeq\section{Concluding remarks}

We have argued that for any choice of $m$ and the central charge $c$, the object $\Omega_n^{(m)}$ of eq. (\ref{hmli}) can be interpreted as an $n$-point correlation function in a conformal field theory with a chiral symmetry algebra generated by the fields $T,J^3,J^-$ with OPEs (\ref{tt}-\ref{jzjw}). This conformal field theory is solvable in the sense that its correlation functions are known in terms of Liouville correlation functions. Liouville theory itself has been solved in the sense that its three-point function on the sphere is explicitly known, while its $n$-point functions on arbitrary Riemann surfaces can in principle be deduced thanks to the conformal symmetry of the theory. (See \cite{tes01a} for a review.) The solution of the new theory is therefore not very explicit, and no closed formula can be written for its three-point function except for some ``degenerate'' values of $m$ (\ref{mpq}) when it satisfies a differential equation.

This differential equation provided the most non-trivial test of our claims. In the case $m=2$, we compared the differential equation deduced from our chiral symmetry algebra with the third-order BPZ equation, and found explicit agreement. We expect such an agreement to hold for all degenerate values of $m$, and it would be interesting to check this beyond case-by-case calculations. 

There may exist other theories based on the same symmetry algebra. For instance, in the case $m=1$, the non-rational, non-unitary $\Hp$ model shares its symmetry algebra with the unitary $AdS_3$ WZW model and with the rational $SU_2$ WZW models. 
For general values of $m$, our model remains non-rational and non-unitary. (The lack of unitarity still follows from Gawedzki's $\Hp$-model argument \cite{gaw91}, although it is not quite clear what the scalar product on the spectrum is, as $J^-$ has no conjugate field.) It would be interesting to construct rational or unitary theories based on the same symmetry algebra. Note however that our reasoning around eq. (\ref{dp}), which showed that in the case $m=1$ the modes of $T,J^-$ and $J^3$ did generate the whole $\Hp$ spectrum, relied on the fact that the spectrum is purely continuous. But the $AdS_3$ and $SU_2$ WZW models involve discrete and finite-dimensional representations respectively, which presumably cannot be generated from the modes of $T,J^-$ and $J^3$. Therefore, a rational theory based on the $(T,J^3,J^-)$ algebra would probably differ from the $SU_2$ WZW model and have a smaller spectrum. 

We have not assumed any restrictions on the choice of the parametrs $m$ and $c$. For some purposes it might be useful to restrict say the central charge of the theory and the conformal dimensions of the fields to be real. Such restrictions could be dictated by particular applications. 
From the point of view of conformal field theory, the new theories are however well-defined, and their correlators manifestly satisfy crossing symmetry, a very stringent constraint. It is therefore possible to study issues like the solution of these theories on Riemann surfaces with boundaries.

\acknowledgments{I am grateful to Volker Schomerus, Thomas Quella and Vladimir Fateev for useful discussions and comments on the manuscript, and to DESY, Hamburg for hospitality while part of this work was done. In addition, I thank the theoretical physics groups at King's College London and Sogang University in Seoul for listening and reacting to informal talks on this subject.}

%\bibliographystyle{JHEP-2}
%\bibliography{992}

\end{document}